# Evolution of the Tohoku earthquake aftershocks in the framework of the phenomenological theory of aftershocks

O.D. Zotov[1,2], A.V. Guglielmi[1]

[1]*Schmidt Institute of Physics of the Earth, Russian Academy of Sciences; Bol'shaya Gruzinskaya str., 10, bld. 1, Moscow, 123242 Russia;*

[2]*Borok Geophysical Observatory of the Schmidt Institute of Physics of the Earth, Russian Academy of Sciences, 142, Borok 152742, Russia*

ozotov@inbox.ru (O.Z.), guglielmi@mail.ru (A.G.)

**Abstract**

The aftershocks of the Tohoku earthquake are analyzed in light of the phenomenological theory of aftershocks. The theory is based on the concept of an earthquake source as a dynamic system, the state of which is described by a deactivation coefficient. The concept of the proper time of the source is used, which in general differs from world time. The "underground clock" and the clock showing world time have been synchronized. For an observer using the underground impacts as time markers, the flow of world time will be uneven. The admissibility and effectiveness of the idea of this specific relativity of time is shown. Three phases of relaxation of the source after the main shock were discovered. In the initial phase, the deactivation coefficient is zero. In the main phase, the deactivation coefficient has a finite value and does not change over time. In the recovery phase, the deactivation coefficient changes randomly over time. The sharp transitions between phases resemble the phenomenon of bifurcation.

## 1. Introduction

The Tohoku earthquake occurred on March 11, 2011, off the east coast of Honshu Island, Japan [1]. The magnitude of the main shock was $M = 9.1$, the depth of the hypocenter was 29 km. After the main shock, aftershocks continued for a long time. The magnitude of the strongest aftershock was $M = 7.9$, which is consistent with Bath's law [2]. The frequency of aftershocks $n(t)$, while experiencing some fluctuations, has on average decreased over time. Figure 1 shows the variation in daily frequency over 200 days after the mainshock. The graph is based on data from the USGS/NEIC catalog [https://earthquake.usgs.gov/earthquakes/search/]. The total number of aftershocks during the specified time period was 4,537.

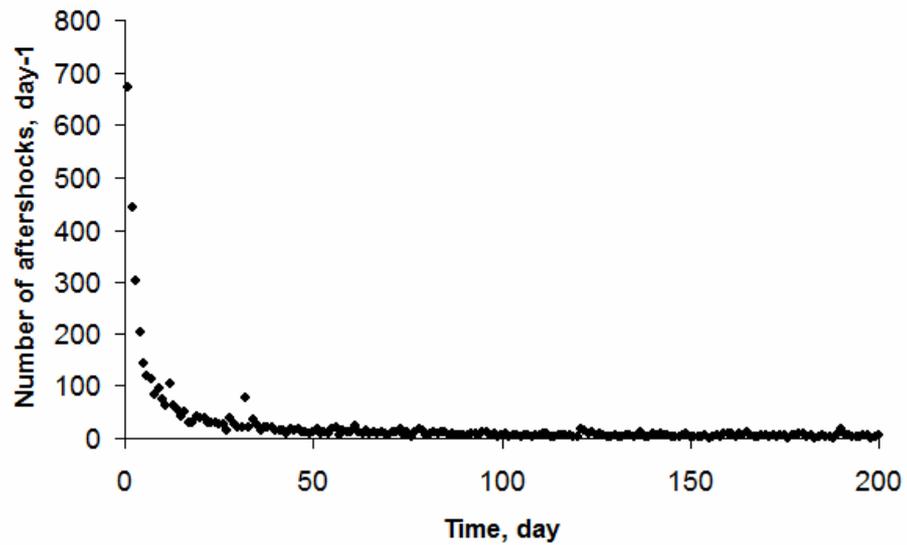

**Fig. 1**. Aftershock frequency versus time after the mainshock of the Tohoku earthquake.

We performed a preliminary analysis of the data presented in Figure 1 using the Hirano-Utsu fitting formula $n(t) = k/(c+t)^p$ [3, 4]. It turned out that the parameter $p = 1.2$ with the determination coefficient $R^2 = 0.98$. We then approximated the observed data using the formula $n(t) = k/(c+t)$, known as Omori's law [5]. The coefficient of determination also turned out to be high: $R^2 = 0.98$.

The purpose of this work is to provide a more detailed analysis of the Tohoku earthquake aftershocks using experimental research methods developed within the framework of the phenomenological theory of aftershocks. The theory is briefly presented in the paper [6], as well as in the Appendix to this paper. A detailed description of the theory is given in review papers [7–12].

The theory is based on the idea of a source as a simple dynamic system. The state of the source is described by the deactivation coefficient $\sigma(t)$. The deactivation coefficient is calculated from the aftershock frequency data using the formula

$$\sigma(t) = \frac{d}{dt}\langle g(t) \rangle. \qquad (1)$$

Here $g(t) = 1/n(t)$ is an auxiliary function, and the angle brackets denote optimal smoothing of the auxiliary function. We imagine the source as a "black box without an entrance". The aftershock flow is considered as an output signal, from which we need to extract information about the dynamics of the source by studying the variation of the deactivation coefficient $\sigma(t)$. For example, if $\sigma = \text{const}$, then the relaxation of the source occurs according to Omori's law. If a monotonically increasing or monotonically decreasing function is a function of time, then relaxation occurs according to the Hirano-Utsu law, with $p > 1$ or $p < 1$, respectively.

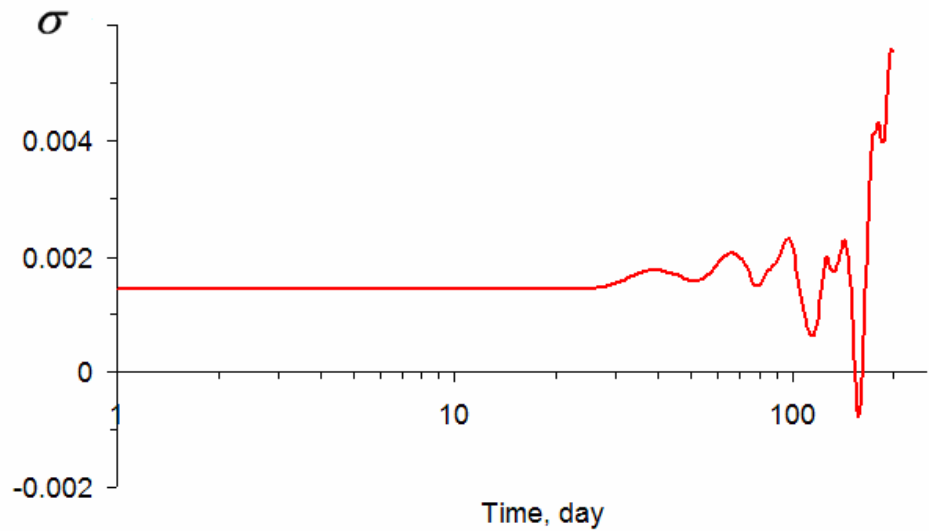

**Fig. 2**. Variation of the deactivation coefficient of the Tohoku earthquake source.

Let us consider Figure 2, which shows the evolution of $\sigma(t)$ for the Tohoku earthquake. We see a non-monotonic dependence of the deactivation coefficient on time. Thus, the Hirano-Utsu law is not satisfied. As for the Omori law, it is fulfilled, but only at the first stage of evolution, which we call the Omori epoch. In the Omori epoch, evolution proceeds in a completely predictable manner. The duration of the epoch is 30 days, the deactivation coefficient is $\sigma = 0.0014$.

The first stage ends with a bifurcation, after which the second stage of chaotic, unpredictable behavior of the deactivation coefficient begins. Thus, Omori's law is generally valid, but it is not holistic.

It is convenient and productive to use the concept of the proper time of the source $\tau$, which in general differs from the world time $t$. Let's determine the proper time using the formula

$$\tau = \int_0^t \sigma(t')dt'. \qquad (2)$$

In this case, the equation of aftershock evolution will take the form of the simplest nonlinear differential equation

$$\frac{dn}{d\tau} + n^2 = 0. \qquad (3)$$

From this we obtain the holistic law of aftershocks

$$n(t) = \frac{n_0}{1 + n_0 \tau(t)}. \qquad (4)$$

Here $n_0 = n(0)$ is the initial condition. Law (4) differs from Omori's law only in that in it the world time is replaced by the proper time of the source.

We see that the tools of the phenomenological theory of aftershocks enrich the arsenal of means suitable for the experimental study of earthquakes. In this work we focused on the source proper time. The abstract idea of an "underground clock" whose movement does not always coincide with the movement of world time allows for options to be chosen. We will present an unexpected way to count the proper time. It will give us a chance to see the aftershocks of the Tohoku earthquake in an unexpected light.

## 2. Proper time of the earthquake source

Our idea is to look at the aftershock sequence from two perspectives. On the one hand, and we have already talked about this, we consider aftershocks as signals that contain information about the earthquake source as a certain dynamic system. The second look may seem strange. We propose to look at the ordered sequence of aftershocks as markers of the proper time of the earthquake source. In other words, at the moment when an aftershock occurs, we say that a unit of proper time has passed since the previous aftershock.

Let's number our aftershocks with numbers $k$ = 1, 2, 3, .., 4537. Each value of $k$ corresponds to a world time $t_k$ of aftershock excitation, as given in the USGS/NEIC catalog. Let us introduce the abstract concept of continuous dimensionless proper time $x$ (from the word "хронос", or χρόνος). It is obvious that a countable set $k$ is a subset of an abstract set $x$, which has the cardinality of the continuum.

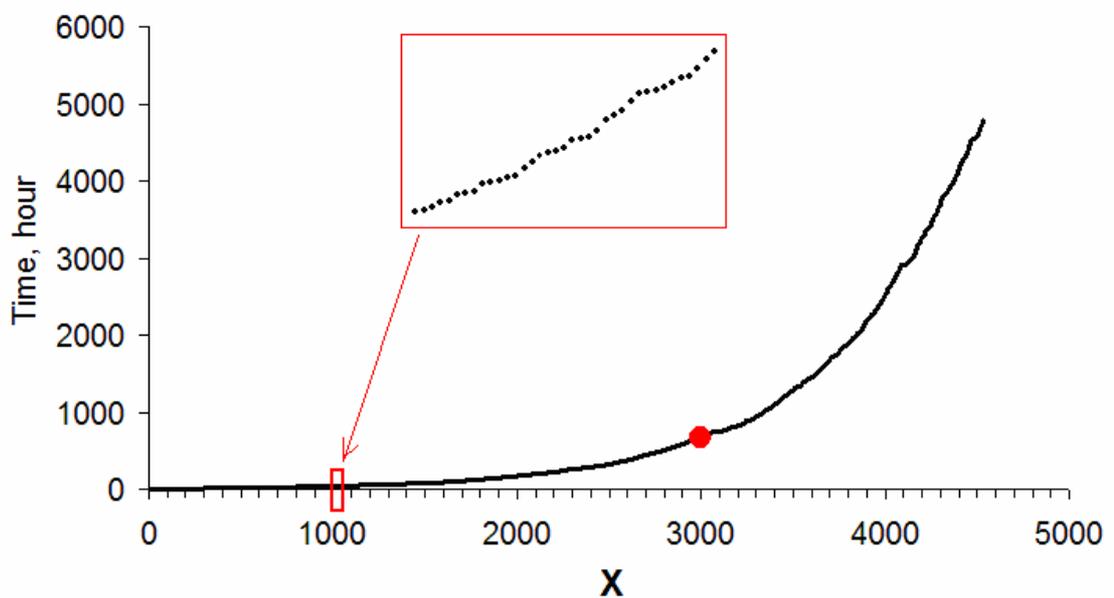

**Fig. 3**. Dependence of the time of occurrence on the aftershock number of the Tohoku earthquake. The points on the coordinate plane ($x$, $t$) represent the excitation times of 4,537 aftershocks. The red dot marks the end of the Omori epoch. A special window shows a fragment of the set of points on an enlarged scale.

Our next step is to find a function $t(x)$ by which we can synchronize our "underground clock" with the clock showing world time. To do this, we plot points $t_k$ on the coordinate plane ($x$, $t$). The result, shown in Figure 3, exceeded our wildest expectations. Remarkably, the dots lined up along a very smooth curve, reminiscent of an exponential function. Approximation of points using the function $t(x) = a\exp(bx)$ gives values of $a = 7.2$, $b = 0..0014$ with a determination coefficient of $R^2 = 0.97$. Obviously, for an observer using underground impacts as time markers, the flow of world time will be uneven.

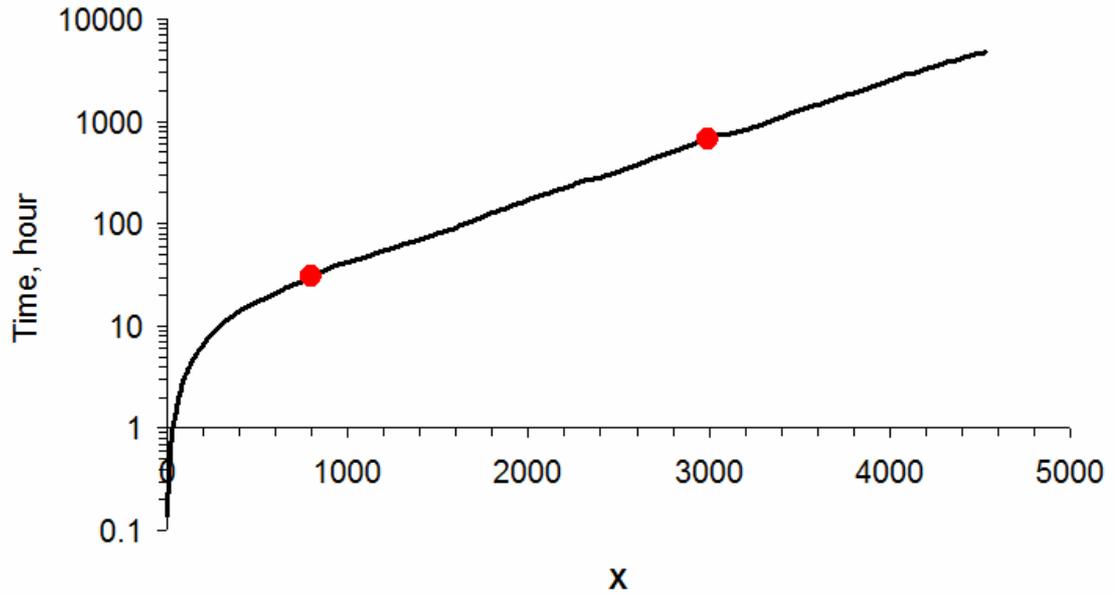

**Fig. 4**. Same as Figure 3, but on a semi-log scale. The left red dot divides a clearly heterogeneous set of points into two homogeneous subsets.

Figure 4 gives us an idea that the experimental data set we selected for our study is clearly heterogeneous. The first 800 or so aftershocks and subsequent aftershocks reflect two significantly different relaxation regimes of the Tohoku earthquake source. It would be possible to select a fitting function, for example, take a function of the form $t(x) = ax \exp(bx)$, which would approximate holistically all experimental points, but we will do otherwise. We will analyze separately two subsets of experimental points, separated in Figure 4 by the left red dot.

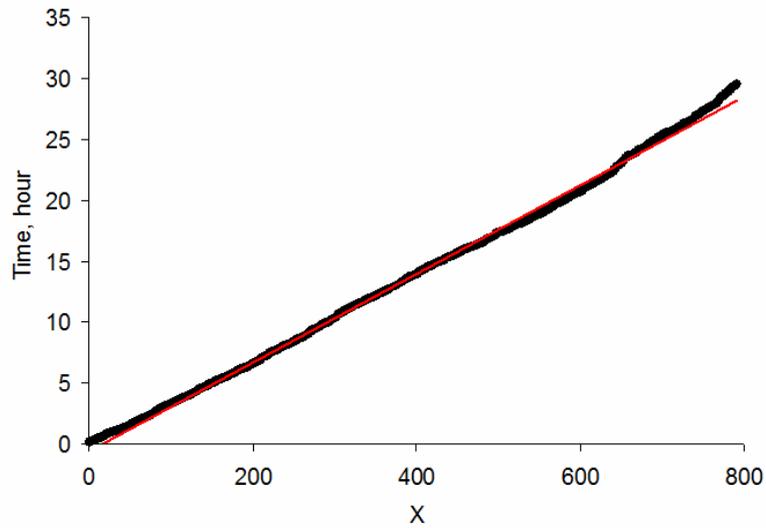

**Fig. 5**. Approximation of a subset of points $k = 1, 2, .., 800$.

The first homogeneous array is shown in Figure 5. Experimental points (black line) are approximated by the red line

$$t(x) = \alpha x. \qquad (5)$$

Here it is $\alpha = 0.036$ hours, and the determination coefficient is $R^2 = 0.99$.

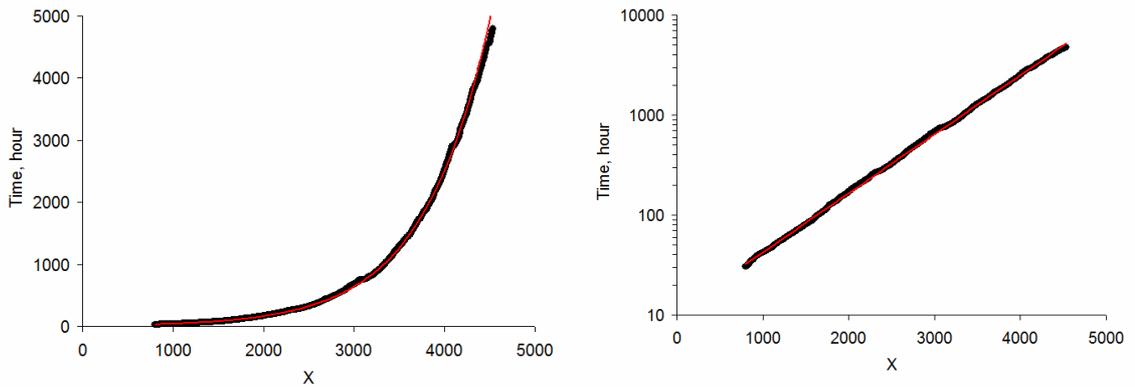

**Fig. 6**. Approximation of a subset of points $k = 801, 802, .., 4537$.

The second homogeneous array is shown in Figure 5 in two views. The experimental points are well approximated by an exponential function.

$$t(x) = \beta \exp(\gamma x). \qquad (6)$$

Here $\beta = 10.8$, $\gamma = 0.0014$. The coefficient of determination is $R^2 = 0.99$.

The initial relaxation phase of the source lasted only 30 hours (1.25 days), but during this time 800 aftershocks occurred. We use the synchronization function (5) to calculate the average universal time interval between adjacent aftershocks

$$T = \frac{dt}{dx} = \alpha = 130 \text{ c.} \qquad (7)$$

The constancy of $T$ in the first phase of evolution contrasts sharply with the subsequent progressive increase in the average time between adjacent aftershocks: $T(x) = \beta\gamma \exp(\gamma x)$. A plausible assumption suggests itself about the phase transition of the state of the Tohoku earthquake source 30 hours after the main shock.

The state of the focus within the framework of the phenomenological theory [6] we describe by the deactivation coefficient

$$\sigma = \frac{d}{dx} \ln T. \qquad (8)$$

According to formula (7) we find that $\sigma = 0$ in the first phase of the evolution of the source. During the phase transition, the deactivation coefficient jumps to a finite value. Indeed, taking into account formula $T = dt/dx$ we find $\sigma = \gamma = 0.0014$. We would like to draw special attention to the fact that we used a very unusual chronometry of aftershocks. Nevertheless, we obtained a deactivation coefficient that coincides with that previously calculated for the Omori epoch when ordering the aftershocks by universal time (see Figure 2).

We already know that the Omori epoch ends with a bifurcation, after which the evolution of the source proceeds in an unpredictable manner. The second phase transition occurs differently than the first. If during the first phase transition there is a jump in the deactivation coefficient, then during the second transition the first derivative with respect to time of the bifurcation coefficient $\theta = d\sigma/dx$ changes abruptly. We will call the time interval between two bifurcations the main phase of relaxation of the source. The last phase, the phase of chaotic behavior of the deactivation coefficient, can naturally be called the recovery phase.

Let's sum it up. Three phases of Tohoku earthquake aftershock evolution have been identified. In the initial phase, the deactivation coefficient is zero. In the main phase, the deactivation coefficient has a finite positive value and does not change over time. In the recovery phase, the deactivation coefficient changes randomly over time. The transitions between phases are quite abrupt. They resemble the phenomenon of bifurcation. Bifurcations indicate a change in the physical properties of the source, vaguely reminiscent of the well-known first- and second-order phase transitions in physics in the case of the first and second bifurcations, respectively.

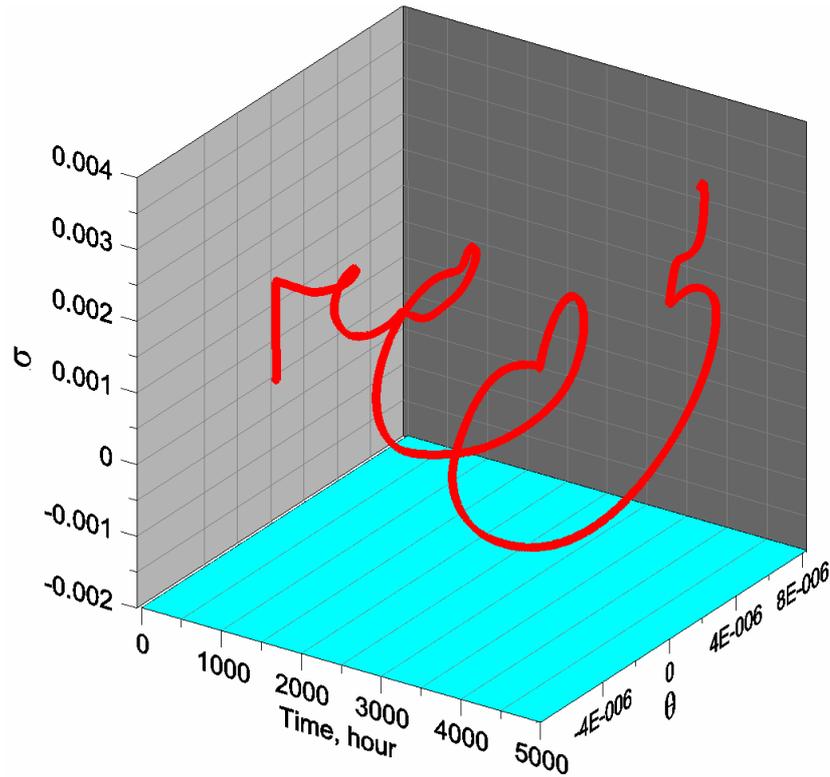

**Fig. 7**. Phase trajectory of the Tohoku earthquake source.

In conclusion of this section of the paper, we present for clarity a phase trajectory of the source in the extended phase space ($\sigma, \theta, t$). Here $\theta = d\sigma/dt$.

### 3. Discussion

The mathematical foundations of aftershock physics were developed by Omori at the end of the century before last and continued by Hirano and Utsu in the last century. At the beginning of the current century, a phenomenological theory of aftershocks was developed. The accumulated collective experience allows us to point out a number of theoretical approaches to the processing and analysis of aftershocks. Omori's law makes it possible to extract from the observation of aftershocks one number, namely the parameter *k*, which quantitatively characterizes a given specific earthquake source. The Hirano-Utsu method allows one to determine two numbers, *k* and *p*. The phenomenological theory makes it possible to additionally characterize the source by the function $\sigma(t)$.

Each of the developed methods has its own advantages and limitations. In the context of this work, we will be interested in comparing two methods for calculating the deactivation coefficient. They differ from each other in the way they time aftershocks. In the continuous approach to the aftershock flow, synchronization of proper and world times is carried out using function $\tau(t)$, and in the discrete approach, using function $x(t)$. Both methods are consistent with each other [6], but differ in their technical procedures. With both approaches we obtain the same values of the deactivation coefficient in the Omori epoch. However, the continuous approach leads to a reliable conclusion about the bifurcation of the source at the end of the Omori epoch (see Figure 2), while with the discrete approach we do not find any signs of bifurcation. On the other hand, the discrete approach made it possible to identify the initial phase of the evolution, in which the deactivation coefficient is zero (see Figure 4). The transition to the main phase occurs abruptly. We could not see this bifurcation in Figure 2 for an obvious reason. The first phase was short-lived, but during its continuation, 800 aftershocks occurred after the main shock of the Tohoku earthquake.

Let us now discuss the issue of the uneven flow of time. It is natural to accept the course of world time as uniform. Then the course of the proper time *x*(*t*) will be uniform in the initial phase of the evolution of the source and non-uniform in the main and recovery phases. In our opinion, the opposite statement also has a certain meaning. Overcoming the natural attitude can be helpful [13]. An example is the detection of foreshock convergence and aftershock divergence when ordering tremors by proper time [14].

We assume that the unevenness of the flow of proper time according to world time clocks implicitly indicates the non-stationarity of the parameters of source. The deactivation coefficient characterizes the functioning of the source in general, but does not contain information about its parameters. When parameters change during operation, the rate of aftershock excitation can change in the same way as the period of oscillation of a pendulum changes when the length of the suspension changes, or when the gravitational force acting on the pendulum changes. However, this kind of reasoning takes us beyond the phenomenological theory of aftershocks.

## 4. Conclusion

The study of aftershocks is of exceptional interest in itself, but the real significance of experimental aftershock research is revealed when the problem of extracting information about the source from data on the evolution of aftershocks is posed. To solve the problem, we used the phenomenological theory that describes the relaxation of the source after the main shock. We focused on the specific example of the Tohoku earthquake aftershocks. To describe the evolution of aftershocks, the proper time of the source is used, which in general differs from universal time. Two methods have been developed for synchronizing the "underground clock" and the clock that keeps track of world time.

The main result of our work is that we were able to identify three phases in the dynamics of the source after the main shock – the initial phase, the main phase and the recovery phase. At the boundaries between phases, bifurcations are observed, reminiscent of phase transitions in physical kinetics. The external manifestation of the three-phase dynamics of the source is that in the initial phase the deactivation coefficient $\sigma(t)$ is zero, in the main phase $\sigma = \text{const}$, and in the recovery phase $\sigma(t)$ experiences chaotic unpredictable variations. At the first bifurcation, a jump in the deactivation coefficient is observed. At the second bifurcation, the function $\sigma(t)$ remains continuous, but its derivative changes abruptly.

*Acknowledgments*. We express our deep gratitude to our colleagues B.I. Klain and A.D. Zavyalov for their long-term cooperation, during which a phenomenological theory of aftershocks was developed. We thank colleagues at the US Geological Survey for lending us their earthquake catalogs USGS/NEIC for use.

The work was carried out according to the plan of state assignments of the Institute of Physics of the Earth of the Russian Academy of Sciences.

**Appendix**

Two representations of the law of aftershock evolution within the framework of phenomenological theory

1. Hyperbolic decay of aftershock frequency with proper time $\tau = \int_0^t \sigma(t') dt'$:

$$\frac{dn}{d\tau} + n^2 = 0, \quad n(t) = \frac{n_0}{1 + n_0 \tau(t)}$$

2. Exponential decay of aftershock frequency with proper time $x = x(t)$:

$$\frac{dn}{dx} + \sigma n = 0, \quad n(x) = n_0 \exp\left(-\int_0^x \sigma(x') dx'\right).$$

In the discrete description of aftershocks, two representations of the law sound like this:

1. The average intervals between successive aftershocks form an arithmetic progression.
2. The average intervals between successive aftershocks form a geometric progression.